\newcommand{\vecj}{\mbox{\boldmath$j$}}
\newcommand{\vecA}{\mbox{\boldmath$A$}}

\newcommand{\vecB}{\mbox{\boldmath$B$}}

\newcommand{\vecV}{\mbox{\boldmath$V$}}
\newcommand{\vecL}{\mbox{\boldmath$L$}}
\newcommand{\vece}{\mbox{\boldmath$e$}}

\newcommand{\dfd}{{\rm d}}

\newcommand{\half}{\frac{1}{2}}
\newcommand{\vecr}{\mbox{\boldmath$r$}}
\newcommand{\vecv}{\mbox{\boldmath$v$}}
\newcommand{\vecp}{\mbox{\boldmath$p$}}
\newcommand{\vecomega}{\mbox{\boldmath$\omega$}}
\newcommand{\vecOmega}{\mbox{\boldmath$\Omega$}}

\documentclass[aps,pre,preprint,groupedaddress]{revtex4}
\usepackage{graphicx}

\begin{document}


\title{Magnetic dynamics of simple collective modes in a two-sphere plasma model}

\author{Hanno Ess\'en}
\affiliation{Department of Mechanics, KTH\\ SE-100 44 Stockholm,
Sweden}


\date{2005 November}

\begin{abstract}
A plasma blob is modeled as consisting of two homogeneous spheres
of equal radius and equal but opposite charge densities that can
move relative to each other. Relative translational and rotational
motion are considered separately. Magnetic effects from the
current density caused by the relative motion are included.
Magnetic interaction is seen to cause an inductive inertia. In the
relative translation case the Coulomb attraction, approximately a
linear force for small amplitudes, causes an oscillation. For a
large number of particles the corresponding oscillation frequency
will not be the Langmuir plasma frequency, because of the large
inductive inertia. For rotation an external magnetic field is
included and the energy and diamagnetism of the plasma in the
model is calculated. Finally it is noted how the neglect of
resistivity is motivated by the results.
\end{abstract}


\maketitle


\section{Introduction}
In order to study qualitative effects of magnetism in plasma
dynamics a very simple model is introduced. Two overlapping
homogeneous spheres of equal radii, and of equal but oppositely
signed charge densities, are assumed to move, relative to each
other, with negligible dissipation (resistivity) under the
influence of electric and magnetic interaction. The neglect of
dissipation is motivated at the end. We first treat relative
translation (oscillation) and then relative rotation.

Many years ago Tonks and Langmuir \cite{tonks} carefully derived
an equation of motion for a collective translational motion of the
electrons relative to the positive ions. Their derivation seems to
indicate that there should be a universal frequency for this mode,
\begin{equation}\label{eq.plasma.freq}
\omega_{\rm p} =\sqrt{\frac{n e^2}{m}},
\end{equation}
the plasma frequency, depending only on the electron number
density $n$. If a large number of electrons move relative to the
positive ions one gets a large current and thus it seems as if
magnetic effects should affect the result. Bohm and Pines
\cite{bohm} studied the influence of magnetic interaction on
plasma modes but they did not come up with any explicit correction
to $\omega_{\rm p}$. In the textbook by Goldston and Rutherford
\cite{goldston} the absence of magnetic effects is said to be due
to a displacement current that compensates for the electron
current. In this article the problem is approached from a very
fundamental starting point: the relevant Lagrangian density. The
conclusion is that the frequency is lowered by the large inductive
inertia [see Eq. (\ref{eq.freq.sq})].

We then study the relative rotation of the two charged spheres. If
magnetic interaction is neglected the kinetic energy is simply
determined by angular momentum and moment of inertia. When
magnetic interaction is included the kinetic energy for a given
angular momentum is much smaller. The reason for this is again
that the effective moment of inertia will be dominated by
inductive inertia. By adding an external magnetic field to the
model we can calculate the response of our model plasma and it
turns out to be diamagnetic.

\section{Separation of overall translation and rotation}
The kinetic energy of any system of particles
\begin{equation}\label{eq.kin.energy.part}
T = \sum_i \half m_i \vecv_i^2
\end{equation}
can be written
\begin{equation}\label{eq.sep.collect.transl.rot}
T=\half M \vecV^2 +\half \vecOmega {\sf J}\vecOmega +T'
\end{equation}
where $M$ is total mass, ${\sf J}$ is the (instantaneous) inertia
tensor, $\vecV$ center of mass velocity, and $\vecOmega$ is a well
defined average angular velocity \cite{jellinek,essen93}. $\vecV$
is chosen so that $\vecp=M\vecV$ is the total momentum of the
system and $\vecOmega$ so that $\vecL={\sf J}\vecOmega$ is the
total (center of mass) angular momentum. $T'$ is the kinetic
energy of the particles relative to the system that moves with the
center of mass velocity and rotates with the average angular
velocity. We call this the co-moving system. One can introduce
generalized coordinates so that there are six degrees freedom
describing center of mass position and of average angular
orientation, while $T'$ depends on the remaining $3N-6$
generalized coordinates.

Now consider a blob of plasma that consists of two spheres of
particles, one of positively, and one of negatively charged
particles. For spheres the inertia tensor ${\sf J}$ can be
replaced by a single moment of inertia $J$. The total kinetic
energy $T$ is the sum of the kinetic energy, $T_1$, of the
positive particles, and of the kinetic energy, $T_2$, of the
negative charges.

We first perform the transformation above to the co-moving systems
separately for the positive and the negative particles. This
gives,
\begin{equation}\label{eq.pos.and.neg.collective}
T =\half M_1 \vecV_1^2 +\half M_2 \vecV_2^2 +\half J_1
\vecOmega_1^2 +\half J_2 \vecOmega_2^2 +T'_1 +T'_2 ,
\end{equation}
for the total kinetic energy, see Fig.\ \ref{FIG0}. In a second
step we then introduce the co-moving system for the total system.
We thus introduce,
\begin{eqnarray}
M=M_1+M_2 ,  & \mbox{\hskip 0.5cm} & J = J_1 +J_2 ,\\
\mu =M_1 M_2/M , & \mbox{\hskip 0.5cm} & I = J_1 J_2/J ,
\end{eqnarray}
total mass and reduced mass as well as total moment of inertia and
reduced moment of inertia. In terms of these one finds,
\begin{eqnarray}
\label{eq.v.transf.1} \vecV = (M_1\vecV_1 +M_2\vecV_2)/M &
\mbox{\hskip 0.5cm}
& \vecV_1 = \vecV +M_2 \vecv /M \\
\label{eq.v.transf.2} \vecv =\vecV_1 -\vecV_2 & \mbox{\hskip
0.5cm} & \vecV_2 = \vecV -M_1 \vecv /M ,
\end{eqnarray}
for the total center of mass velocity $\vecV$, and relative
velocity $\vecv$, of the two spheres. Finally,
\begin{eqnarray}
\vecOmega = (J_1\vecOmega_1 +J_2\vecOmega_2)/J & \mbox{\hskip
0.5cm}
& \vecOmega_1 = \vecOmega +J_1 \vecomega /J \\
\vecomega =\vecOmega_1 -\vecOmega_2 & \mbox{\hskip 0.5cm} &
\vecOmega_2 = \vecOmega - J_2 \vecomega /J ,
\end{eqnarray}
gives the total average angular velocity $\vecOmega$, and the
relative angular velocity $\vecomega$, of the two oppositely
charged spheres. In terms of these quantities we get the
expression,
\begin{equation}\label{eq.collective.transformed}
T =\half M \vecV^2 +\half \mu \vecv^2 +\half J \vecOmega^2 +\half
I \vecomega^2 +T',
\end{equation}
for the total kinetic energy of our two-sphere system. The degrees
of freedom in $T'$ are assumed to be random and not to produce any
net charge or current density. They will be ignored henceforth.

\section{Lagrangian including magnetic interactions}
Maxwell's equations and the equations of motion for the charged
particles with the Lorentz force, can all be derived from a single
Lagrangian via the variational principle \cite{landau2}. The
Lagrangian has three parts, particle, interaction, and field
contributions. If radiation is neglected the field does not have
independent degrees of freedom, but is determined by particle
positions and velocities. Using the non-relativistic form for the
kinetic energy one then gets,
\begin{equation}
\label{eq.LtotNoRad2} L = \sum_i \left(\frac{1}{2} m_i\vecv_i^2 +
\frac{q_i}{2c} \vecv_i \cdot \vecA(\vecr_i) -
\frac{q_i}{2}\phi(\vecr_i) \right) ,
\end{equation}
where,
\begin{equation}
\label{eq.coul.pot}
 \phi(\vecr,t) = \sum_i
\frac{q_i}{|\vecr -\vecr_i|},
\end{equation}
and
\begin{equation}
\label{eq.darwin.A.ito.velocity} \vecA(\vecr,t) = \sum_i\frac{q_i
[\vecv_i + (\vecv_i\cdot\vece_i) \vece_i] }{2c|\vecr-\vecr_i|} .
\end{equation}
Here the position and velocity vectors of the particles are
$\vecr_i$ and $\vecv_i$ respectively, $m_i$ and $q_i$ their masses
and charges, and $\vece_i = (\vecr-\vecr_i)/|\vecr-\vecr_i|$
(Darwin \cite{darwin}, Jackson \cite{jackson3}, Schwinger {\it et
al.} \cite{schwingeretal}, Ess\'en \cite{essen96,essen99}). The
vector potential here is in the Coulomb gauge, and this
essentially means that all velocity dependence of the interaction
appears in the magnetic part, leaving the Coulomb interaction
energy in its static form.

When the expressions (\ref{eq.coul.pot}) and
(\ref{eq.darwin.A.ito.velocity}) are inserted into equation
(\ref{eq.LtotNoRad2}) one finds infinite contributions from
self-interactions. When these are discarded, so that each particle
only interacts with the field from the others, one obtains
\begin{equation}
\label{eq.LtotNoRad.darwin} L = \sum_i \frac{1}{2} m_i\vecv_i^2+
\sum_{i<j}\frac{q_i q_j}{r_{ij}} \frac{ [\vecv_i \cdot\vecv_j
+(\vecv_i\cdot\vece_{ij})(\vecv_j\cdot\vece_{ij})]}{2c^2} -
\sum_{i<j}\frac{q_i q_j}{r_{ij}},
\end{equation}
where now $r_{ij}$ is the distance between particles $i$ and $j$
and $\vece_{ij}$ is the unit vector pointing from $i$ to $j$. This
is the so called Darwin Lagrangian \cite{darwin} for the system.
We can write it,
\begin{equation}\label{eq.parts.L.darwin}
L = T + L_{\rm mag} -\Phi,
\end{equation}
where, $T$, is kinetic energy and, $\Phi$, the Coulomb electric
interaction energy. The magnetic part can also be written
\begin{equation}\label{eq.Lmag.in.darwin}
L_{\rm mag} =\sum_i \frac{q_i}{2c} \vecv_i \cdot \vecA(\vecr_i) =
\frac{1}{2c}\int \vecj(r)\cdot \vecA(r) \,\dfd V .
\end{equation}
Here it is important the the vector potential is divergence free
($\nabla\cdot\vecA=0$, Coulomb gauge). The Darwin Lagrangian thus
includes both electric and magnetic interactions between the
particles and is valid in when radiation can be neglected.

\section{Relative translational motion}
The Coulomb interaction, $\Phi$, between two overlapping charged
spheres is calculated in Appendix \ref{app.coul}. The magnetic
interaction between two charged spheres in relative translational
motion is calculated in Appendix \ref{app.mag.transl} for the case
of small displacement of centers of the spheres ($r \ll R$).
Keeping only the quadratic term, one finds,
\begin{equation}\label{eq.lagr.transl}
L_{\rm rel}= \half \mu \vecv^2+ \frac{4 Q^2}{10 R c^2} \vecv^2
-\half \frac{Q^2}{R^3} \vecr^2.
\end{equation}
Here $\vecr$ is the vector to center of the positive sphere from
the center of the negative, so that $\dot{\vecr}=\vecv$. The
center of mass motion decouples, and we assume that the random
motions decouple. This is then the relevant Lagrangian for the
relative collective translation. It can be written,
\begin{equation}\label{eq.lagr.transl.rel}
L_{\rm rel}= \half {\cal M} \vecv^2 -\half {\cal K} \vecr^2 ,
\end{equation}
where,
\begin{equation}\label{eq.eff.inert.mass}
{\cal M}= \mu + \frac{4 Q^2}{5 c^2 R} \approx N
m\left(1+\frac{4}{5}\frac{Nr_{\rm e}}{R} \right),
\end{equation}
and ${\cal K}=Q^2/R^3$. If we assume a proton-electron plasma we
get $M_1 = N m_{\rm p}$, $M_2 = N m$, $\mu = N  m_{\rm p}m/(m_{\rm
p}+m) \approx  N m$, and $Q^2 = N^2 e^2$.  On the right hand side
we have introduced, $r_{\rm e}\equiv\frac{e^2}{mc^2}$, the
classical electron radius. Note that when $N r_{\rm e}/R \gg 1$
the effective mass ${\cal M}$ is entirely due to inductive
inertia.

Clearly the Lagrangian (\ref{eq.lagr.transl.rel}) corresponds to
an oscillating system with angular frequency $\omega_0
=\sqrt{{\cal K}/{\cal M}}$. For this frequency we get explicitly,
\begin{equation}\label{eq.freq.sq}
\omega_0^2 =\frac{\frac{Ne^2}{R^3 m}}{1+\frac{4}{5}\frac{N r_{\rm
e}}{R}}.
\end{equation}
If we introduce the dimensionless number,
\begin{equation}\label{eq.NreoverR}
\nu \equiv N r_{\rm e}/R,
\end{equation}
we see that for $\nu \ll 1$, one obtains, essentially, the
Langmuir plasma frequency, $\omega^2_{\rm p}\propto n e^2/m$, see
Eq.\ (\ref{eq.plasma.freq}). If we reexpress the plasma frequency
in terms of the classical electron radius \cite{hershberger}, Eq.\
(\ref{eq.freq.sq}) can be written in the form,
\begin{equation} \label{eq.re.omega0}
\omega_0^2 = \frac{\nu}{1+\frac{4}{5}\nu} \frac{c^2}{R^2}.
\end{equation}
Thus, when the number of particles is large enough so that $\nu
\gg 1$, this gives,
\begin{equation}\label{eq.largeN.freq}
\omega_0^2= \frac{5 c^2}{4 R^2}.
\end{equation}
For this case the frequency turns out to depend on the size
(radius) of the sphere, but not on the density.

\section{Relative rotational motion}
We now study pure rotational motion of the two charged spheres,
about their coinciding centers of mass, but we include
interaction,
\begin{equation}\label{eq.Le.ext.mag.field}
L_{\rm e} =\frac{1}{c}\int \vecj(\vecr)\cdot \vecA_{\rm e}(\vecr)
\,\dfd V,
\end{equation}
with a constant external magnetic field
$\vecB=\nabla\times\vecA_{\rm e}$. Here $\vecA_{\rm e}=\half
\vecB\times\vecr$ is the vector potential of the external field.
Starting from (\ref{eq.collective.transformed}) and
(\ref{eq.parts.L.darwin}) we find that,
\begin{equation}\label{eq.lagr.rot}
L_{\rm rot}= \half I \vecomega^2  +L_{\rm mag} +L_{\rm e},
\end{equation}
is the relevant Lagrangian for collective relative rotation.

The explicit calculations are sketched in Appendix
\ref{app.mag.int.rot}. One finds that $L_{\rm mag}\sim
\vecomega^2$ and that this term therefore contributes to the
effective moment of inertia, just as it contributed to the
effective mass, ${\cal M}$, in the translational case. The result
can be written,
\begin{equation}\label{eq.rot.lagrange.expl}
L_{\rm rot}=\half {\cal I}\vecomega^2 + \frac{QR^2}{10
c}\vecomega\cdot\vecB,
\end{equation}
where,
\begin{equation}\label{eq.eff.mom.of.inert}
{\cal I}=I+\frac{2}{35}\frac{Q^2 R}{c^2} =
\frac{2}{5}NmR^2\left(1+\frac{1}{7}\frac{Nr_{\rm e}}{R}\right).
\end{equation}
We see that when $\nu =N r_{\rm e}/R \gg 1$ we can neglect the
contribution from mass to the effective moment of inertia ${\cal
I}$. In this limit therefore,
\begin{equation}\label{eq.eff.mom.of.inert.induc}
{\cal I}\approx \frac{2}{35}\frac{Q^2 R}{c^2},
\end{equation}
and there is essentially only inductive moment of inertia. We
assume this below.

\section{Plasma energy and diamagnetism}
In order to investigate the equation of motion we put
$\vecomega=\dot\varphi\vece_z$ and
$\vecB=B(\sin\theta\vece_x+\cos\theta\vece_z)$. The Lagrangian
then becomes
\begin{equation}\label{eq.rot.lagrange.expl.z}
L_{\rm rot}=\half {\cal I}\dot\varphi^2 + \frac{QR^2}{10
c}\dot\varphi\, B\cos\theta.
\end{equation}
In general when, $\partial L/\partial \varphi =0$, one finds that,
$\dot p_{\varphi} =(\dfd/\dfd t)(\partial L/\partial
\dot\varphi)=0$. In our case this gives,
\begin{equation}\label{eq.pphi.const}
p_{\varphi} =\frac{\partial L_{\rm rot}}{\partial \dot\varphi}
={\cal I}\dot\varphi+\frac{QR^2}{10 c} B \cos\theta=\mbox{const.}
\end{equation}
If we assume that $\dot\varphi(t=0) =0$ when $B(t=0) = 0$ we find
that the constant is zero: $p_{\varphi}=0$. At all times we then
find the relation,
\begin{equation}\label{eq.ang.vel.and.B.rel}
 \dot\varphi(t)= \frac{7Rc}{4Q} B(t) \cos\theta,
\end{equation}
between angular velocity and the magnetic field.

To get the energy from $L(\varphi,\dot\varphi)$ one calculates the
Hamiltonian $H=p_{\varphi}\dot\varphi - L$. For the $L_{\rm rot}$
of Eq.\ (\ref{eq.rot.lagrange.expl.z}) one finds,
\begin{equation}\label{eq.hamilt.pphi.B}
H=\frac{1}{2 {\cal I}} \left(p_{\varphi} - \frac{QR^2}{10 c}
B\cos\theta \right)^2.
\end{equation}
Let us consider two special cases of this phase space energy of
the plasma.

We first assume that the external field is zero ($B=0$). The
energy is then given by $E = H = p^2_{\varphi} /2{\cal I}$. Here
$p_{\varphi}$ is the angular momentum of relative rotation. For a
given value of this angular momentum the energy is thus much
smaller when $\nu \gg 1$ than otherwise. This reflects the fact,
repeatedly stressed by the author
\cite{essen96,essen99,essen97,essennordmark}, that for given
momenta the phase space energy of a plasma is lower when there is
net current than in the absence of net current.

Now consider instead the case $p_{\varphi}=0$. For simplicity we
also assume $\theta=0$. One then finds that
\begin{equation}\label{eq.energy.of.B}
H(p_{\varphi}=0)=\frac{1}{2 {\cal I}} \left( \frac{QR^2}{10 c} B
\right)^2
\end{equation}
or, equivalently, using (\ref{eq.eff.mom.of.inert.induc}), that
the energy as function of $B$ is given by,
\begin{equation}\label{eq.energy.of.B.2}
E(B)= \frac{7}{80} R^3 B^2 =\frac{21}{40} \left(\frac{4\pi
R^3}{3}\right)
 \frac{B^2}{8\pi}.
\end{equation}
Note that here, $B^2/8\pi$, is the energy density of the field $B$
in our (gaussian) units. The energy is thus seen to grow
quadratically with the applied magnetic field and our plasma
spheres are strongly diamagnetic. Based on more detailed studies
Cole \cite{cole} has also concluded that plasmas are diamagnetic.
In our model plasma diamagnetism is seen to be closely related to
the diamagnetism of superconductors, as discussed by Ess\'en
\cite{essen05}: the external field induces a current that screens
the external field and reduces it inside. In the absence of
resistance this screening current persists.

\section{Plasma resistivity}
Resistivity is completely neglected in the present model. It has
been pointed out by Kulsrud \cite{kulsrud}, in his book on
astrophysical plasmas, that the negligible resistivity of such
plasmas is in fact closely connected with magnetic induction. In
the present treatment magnetic induction appears in the form of an
inductive inertia that appears naturally as the main physical
parameter in the present model. As early as 1933 Frenkel
\cite{frenkel} suggested that superconductivity is due to
inductive inertia. Frenkel also conjectured that inductive inertia
can lower the energy and cause a phase transition. He did not,
however, make his ideas quantitative. The present model system
gives Frenkel's ideas some quantitative backing.

We note that for a collective momentum $p$ involving $N$ particles
the kinetic energy will be $T = p^2/(2mN)$, when magnetic
interaction is neglected, see Eqs.\ (\ref{eq.lagr.transl.rel}
-\ref{eq.eff.inert.mass}). When the effect of magnetic interaction
is included this becomes $T+E_{\rm mag} = p^2/(2Nm[1+4\nu/5])$ and
we find that $T+E_{\rm mag} \ll T$ when $\nu \gg 1$, assuming that
$p$ remains constant. Collective modes are thus much more
favorable thermodynamically when there is net current.

Plasma resistivity is normally treated by studying the scattering
of individual charged particles. Even with this type of treatment
fast electrons become, so called, runaway electrons and experience
no resistance \cite{alfven}. The present model indicates that
resistivity can not be treated as resulting from the scattering of
individual particles, since the collective motion of many charges
leads to a large inductive non-local effect. All this points in
the same direction, namely that plasmas need not be resistive, in
agreement with our model treatment.

If resistivity had to be included the translational oscillation
would become a damped oscillation and any circulating current
would eventually cease, thereby making the diamagnetic response
temporary. A more immediate limitation of our model is probably
the fact that a current would cause pinching and this would lead
to instabilities that deform of the spherical shape. Lynden-Bell
\cite{lyndenbell} has studied the relativistically spinning
charged sphere and finds that charge concentrates near the equator
(as a result of pinching).

\section{Conclusions}
The model treated in this article is not particularly realistic.
Instead it can be motivated as the simplest possible model within
which one can study plasma phenomena associated with current,
induction, and magnetic interaction energy, in a meaningful way.
Hopefully it also has some novelty. In the literature one can find
a fair amount of work on the radially oscillating plasma sphere
(see e.g. Barnes and Nebel \cite{barnes}, Park {\it et al.}
\cite{park}), but not the modes treated here.  A numerical study
of a rotating convective plasma sphere, modelling a star, by
Dobler {\it et al.} \cite{dobler} shows how complicated more
realistic models necessarily become.

The two-sphere model studied here is therefore valuable as a
device for gaining insight into some very basic plasma phenomena.
As we have seen the most basic of these is the dominance of
inductive inertia in the effective mass ${\cal M}$ of Eq.\
(\ref{eq.eff.inert.mass}), and the effective moment of inertia
${\cal I}$ of Eq.\ (\ref{eq.eff.mom.of.inert}), when the number
$N$ of participating charged particles is large enough. One notes
that $\nu =Nr_{\rm e}/R \approx 118$ for a typical laboratory
plasma of density $n=10^{20}\,{\rm m}^{-3}$ and of radius $R=
1\,$cm, assuming that all particles contribute to the collective
mode. Finally the model also indicates how this large inductive
inertia influences the energy of the plasma and how a plasma
responds diamagnetically to an external magnetic field.

\newpage
\noindent{\Large\bf Appendices}
\appendix
\section{Electrostatic interaction of two overlapping charged spheres}
\label{app.coul} The electrostatic potential from a spherically
symmetric charge density, $\varrho(r)$, is given by,
\begin{equation} \label{eq.pot.from.sphere}  \phi(r) =  4\pi \left(
\frac{1}{r}\int_{0}^{r}\varrho(r') r'\,^2 \,\dfd r' +
\int_{r}^{\infty} \varrho(r') r'\, \dfd r' \right).
\end{equation}
For a homogeneous charged sphere of total charge $Q$ and radius
$R$  with $\varrho = Q/\frac{4\pi R^3}{3}$, for $r\le R$ and zero
for $r>R$, this gives,
\begin{equation} \label{eq.pot.sphere.expl}
\phi(r) = \left\{ \begin{array}{ll}
\frac{3Q}{2R}-\frac{Qr^2}{2R^3} & \mbox{ for $r< R$}\\
   &   \\
\frac{Q}{r} & \mbox{ for $r\ge R$ ,} \end{array} \right.
\end{equation}
and there is a pure Coulomb potential outside the sphere.

Now consider a second, homogeneously charged, sphere of the same
radius and of total charge $q$. If the distance $r$ between the
centers of the two spheres is greater than $2R$ the interaction
energy is clearly given given by,
\begin{equation}\label{eq.two.spheres.outside}
\Phi(r) = \frac{qQ}{r},\hskip 0.5cm \mbox{for $r>2R$}.
\end{equation}
We will now investigate what this interaction energy is when the
distance is smaller so that the two spheres intersect.

The interaction energy can be written
\begin{equation}\label{eq.interact.energy}
\Phi(r) = \int \varrho_q \phi\,\dfd V
\end{equation}
where the volume integration is over the space occupied by the
second sphere, where its charge density, $\varrho_q=q/\frac{4\pi
R^3}{3}$, is different from zero.

We do the integration using the slicing by concentric spheres
depicted in Fig.\  \ref{FIG1}. The volume element is then given by
\begin{equation}\label{eq.vol.element}
\dfd V =\pi \frac{\rho}{r}[R^2-(r-\rho)^2]\dfd\rho .
\end{equation}
When this is integrated between the limits $\rho=r-R$ and
$\rho=r+R$ one should get the volume of the sphere and indeed and
one easily checks that
\begin{equation}\label{eq.sphere.vol}
\int_{\rho=r-R}^{\rho=r+R} \dfd V(\rho) =\frac{4\pi R^3}{3}.
\end{equation}
To do the integral (\ref{eq.interact.energy}) we have to split the
integration range at $\rho=R$ since the function $\phi$ of
(\ref{eq.pot.sphere.expl}) since it changes character at that
radius. This gives
\begin{equation}\label{eq.interact.energy.ints}
\Phi(r) =\varrho_q \left[
\int_{\rho=r-R}^{\rho=R}\left(\frac{3Q}{2R}-\frac{Q\rho^2}{2R^3}
\right) \dfd V + \int_{\rho=R}^{\rho=r+R}  \frac{Q}{\rho} \dfd V
\right].
\end{equation}
The integrals are elementary and the final result, for $r<2R$, is,
\begin{equation}\label{eq.interact.energy.final}
\Phi(r) =\frac{qQ}{160 R^6}\left(192 R^5 - 80 R^3 r^2+ 30 R^2 r^3
-r^5 \right).
\end{equation}
If we assume small displacements, $r \ll R$, and put $q=-Q$,  we
find that,
\begin{equation}\label{eq.inf.interact}
\Phi(r) \approx -\frac{6Q^2}{5R}+\frac{Q^2 r^2}{2 R^3},
\end{equation}
so the potential is quadratic and there is there is a linear
restoring radial force, $F=-kr$, with force constant, $k=Q^2/R^3$.

\section{Magnetic interaction of two charged spheres with relative
translation} \label{app.mag.transl} Here we calculate $L_{\rm
mag}$, the middle term of Eq.\ (\ref{eq.LtotNoRad.darwin}), for
the two charged spheres with velocities $\vecV_1$ and $\vecV_2$ of
Eqs.\ (\ref{eq.v.transf.1} -- \ref{eq.v.transf.2}). This Darwin
term in the Lagrangian gives explicitly,
\begin{eqnarray}
\label{eq.darw.pos.ball} L_{\rm mag} = \frac{q_1^2}{4c^2} \sum_i
\sum_j \frac{\vecV_1^2 +(\vecV_1 \cdot \vece_{ij})^2}{|\vecr_i -\vecr_j|} +\\
\label{eq.darw.pos.neg.balls} \frac{q_1 q_2}{4c^2} \sum_i \sum_k
\frac{\vecV_1\cdot\vecV_2 +(\vecV_1 \cdot \vece_{ik})
(\vecV_2 \cdot \vece_{ik})}{|\vecr_i -\vecr_k|}+\\
\label{eq.darw.neg.pos.balls} \frac{q_2 q_1}{4c^2} \sum_k \sum_i
\frac{\vecV_2 \cdot\vecV_1 +
(\vecV_2 \cdot \vece_{ki})(\vecV_1 \cdot \vece_{ki})}{|\vecr_k -\vecr_i|} +\\
\label{eq.darw.neg.ball}
  \frac{q_2^2}{4c^2} \sum_k \sum_l
\frac{\vecV_2^2 +(\vecV_2 \cdot \vece_{kl})^2}{|\vecr_k
-\vecr_l|},
\end{eqnarray}
for the magnetic interaction. Here the indices $i,j$ refer to the
positive sphere and $k,l$ to the negative. The two terms
(\ref{eq.darw.pos.neg.balls}) and (\ref{eq.darw.neg.pos.balls})
representing interactions between the two oppositely charged
spheres are clearly equal. If we denote the angle in
(\ref{eq.darw.pos.ball}) between $\vecV_1$ and $\vecr_i -\vecr_j$
by $\theta_{ij}$, and similarly for (\ref{eq.darw.neg.ball}), we
thus get
\begin{eqnarray}
\label{eq.L.balls.self}
 L_{\rm mag} = \frac{q_1^2 V_1^2}{4c^2} \sum_i
\sum_j \frac{1+\cos^2 \theta_{ij}}{|\vecr_i -\vecr_j|} +
\frac{q_2^2 V_2^2}{4c^2} \sum_k \sum_l \frac{1+\cos^2
\theta_{kl}}{|\vecr_k -\vecr_l|}
\\
\label{eq.L.balls.int} +\frac{q_1 q_2}{2c^2} \sum_i \sum_k
\frac{\vecV_1\cdot\vecV_2 +(\vecV_1 \cdot \vece_{ik}) (\vecV_2
\cdot \vece_{ik})}{|\vecr_i -\vecr_k|}.
\end{eqnarray}
According to our assumptions the two double sums in
(\ref{eq.L.balls.self}) both represent the internal interaction
between particles uniformly distributed within a sphere of radius
$R$. They must thus be equal, and if we split the Darwin
Lagrangian into
\begin{equation}\label{eq.L.split}
L_{\rm mag}=L_{\rm self}+L_{\rm int},
\end{equation}
where $L_{\rm self}$ stands for (\ref{eq.L.balls.self}) and
$L_{\rm int}$ for (\ref{eq.L.balls.int}), we find
\begin{equation}\label{eq.L.balls.self.2}
L_{\rm self}=\left(\frac{q_1^2 V_1^2}{4c^2}+\frac{q_2^2
V_2^2}{4c^2} \right)\sum_i \sum_j \frac{1+\cos^2
\theta_{ij}}{|\vecr_i -\vecr_j|}.
\end{equation}
Without the squared cosines the double sum would simply be the
Coulomb self interaction of a charged sphere, which can be taken
from (\ref{eq.interact.energy.final}) with $r=0$. Since the
directions of $\vecr_i -\vecr_j$ vary over the sphere it seems
reasonable to estimate the effect of the squared cosine by
replacing it with its spherical average
\begin{equation}\label{eq.sphere.aver}
\overline{\cos^2\theta}\equiv \frac{1}{4\pi}\int_{\Omega}
\cos^2\theta\, \dfd\Omega = \frac{1}{3},
\end{equation}
where $\dfd\Omega=\sin\theta\,\dfd\theta\,\dfd\phi$. Also using,
\begin{equation}\label{eq.double.sum.ball.self.int}
\sum_{i=1}^N \sum_{j=1}^N \frac{1}{|\vecr_i -\vecr_j|} =
\frac{6N^2}{5R},
\end{equation}
we get
\begin{equation}\label{eq.L.balls.self.3}
L_{\rm self}=\left(\frac{q_1^2 V_1^2}{4c^2}+\frac{q_2^2
V_2^2}{4c^2} \right)\frac{4}{3}\frac{6}{5}\frac{N^2}{R} =
\frac{2Q^2}{5R c^2}( V_1^2+  V_2^2),
\end{equation}
where, $Nq_1=Q,\; Nq_2=-Q$. We now introduce the transformation
(\ref{eq.v.transf.1})--(\ref{eq.v.transf.2}). If we put
$M_2/M\equiv\alpha$ and $M_1/M\equiv\beta$ we can write it
\begin{equation}\label{eq.alpha.beta.v.tr}
\vecV_1=\vecV+\alpha\vecv,\; \mbox{and}\;
\vecV_2=\vecV-\beta\vecv,
\end{equation}
where $\alpha+\beta=1$. This finally gives
\begin{equation}\label{eq.L.balls.self.4}
L_{\rm self}=\frac{2}{5}\frac{Q^2}{R c^2}[2V^2
+2(\alpha-\beta)\vecV\cdot\vecv+(\alpha^2+\beta^2)v^2],
\end{equation}
for the magnetic self interactions of the spheres.

For the interaction part we put (\ref{eq.alpha.beta.v.tr}) into
(\ref{eq.L.balls.int}) and get,
\begin{eqnarray}\label{eq.ball.interact.trans}
L_{\rm int} =\frac{q_1 q_2}{2c^2}\left[V^2 \sum_{i,k}
\frac{1+\cos^2 \theta_{ik}}{|\vecr_i -\vecr_k|} -\alpha\beta v^2
 \sum_{i,k} \frac{1+\cos^2 \theta'_{ik}}{|\vecr_i
-\vecr_k|} +\right.
\\
\label{eq.V.v.inter} \left. + (\alpha-\beta)\sum_{i,k}
\frac{\vecV\cdot\vecv +
(\vecV\cdot\vece_{ik})(\vecv\cdot\vece_{ik})}{|\vecr_i -\vecr_k|}
\right].
\end{eqnarray}
We now split the vector $\vecv$ into a part parallel to $\vecV$
and a part perpendicular to $\vecV$, to get
$\vecv=\vecv_{V}+\vecv_{\perp}$. The double sum in
(\ref{eq.V.v.inter}) is then
\begin{equation}\label{eq.double.sum}
\sum_{i,k} \frac{\vecV\cdot\vecv +
(\vecV\cdot\vece_{ik})(\vecv\cdot\vece_{ik})}{|\vecr_i -\vecr_k|}
=Vv_V \sum_{i,k}\frac{1+\cos^2 \theta_{ik}}{|\vecr_i -\vecr_k|} +
Vv_{\perp}\sum_{i,k}\frac{\cos\theta_{ik}
\cos\theta''_{ik}}{|\vecr_i -\vecr_k|}.
\end{equation}
In the above expressions, $\theta_{ik}$, $\theta'_{ik}$, and
$\theta''_{ik}$, are the angles between $\vecr_i -\vecr_k$ and
$\vecV$, $\vecv$, and $\vecv_{\perp}$, respectively. The double
sums are now over points distributed in displaced spheres. If we
assume that the displacement $r$ is small compared to the radius
$R$ we can again approximate the effect of the cosines by
spherical averaging (\ref{eq.sphere.aver}). We then find that
\begin{equation}\label{eq.sphre.av.two}
\overline{1+\cos^2\theta}=\frac{4}{3},\; \mbox{and,}\;
\overline{\cos\theta\cos\theta'}=0,
\end{equation}
where $\theta$ and $\theta'$ represent angles with perpendicular
directions.

According to our assumptions we can now  use Eq.\
(\ref{eq.inf.interact}) and put
\begin{equation}\label{eq.sums.approx.displaced}
\sum_{i,k}\frac{1}{|\vecr_i -\vecr_k|} \approx
\frac{6N^2}{5R}-\frac{N^2 r^2}{2 R^3}.
\end{equation}
Finally then we get (note that $Vv_V =\vecV\cdot\vecv$)
\begin{equation}\label{eq.L.balls.int.2}
L_{\rm int}=-\frac{Q^2}{R
c^2}\left(\frac{2}{5}-\frac{1}{6}\frac{r^2}{R^2} \right)[2V^2
+2(\alpha-\beta)\vecV\cdot\vecv-2\alpha\beta v^2],
\end{equation}
for the magnetic interaction of the two spheres.

If we now add the self interaction, Eq.\ (\ref{eq.L.balls.self.4})
to the result (\ref{eq.L.balls.int.2}) we end up with,
\begin{equation}\label{eq.L.mag.transl.tot}
L_{\rm mag}=\frac{4}{10} \frac{Q^2}{R}  \frac{v^2}{c^2} +
\frac{1}{3} \frac{Q^2}{R}\frac{r^2}{R^2}\, \frac{V^2
+(\alpha-\beta)\vecV\cdot\vecv-\alpha\beta v^2}{c^2},
\end{equation}
for the total magnetic interaction part of the Lagrangian. Here we
have used, $\alpha+\beta=1$. Note also that we have assumed that
$r\ll R$ and that certain angular dependencies have been treated
approximately. If we also assume that $v^2 \ll c^2$ the second
term here is negligible compared to the first.

\section{Magnetic interaction of two charged spheres with relative
rotation} \label{app.mag.int.rot}
 If we put $\varrho_1 = Q/\frac{4\pi R^3}{3}$ and
$\varrho_2 = -\varrho_1$, for $r<R$, and zero outside, the two
spheres have current densities,
\begin{equation}\label{eq.curr.dens.sphere.rot}
\vecj_i (\vecr) = \varrho_i \vecOmega_i \times \vecr,\;\; i=1,2.
\end{equation}
Divergence free vector potentials, that match smoothly with dipole
field vector potentials outside the spheres, are then given by
\cite{essen89},
\begin{equation}\label{eq.vec.pot.rot.sphere}
\vecA_i(\vecr) = \frac{2\pi}{5c}\left(\frac{5}{3}R^2-r^2
\right)\vecj_i(\vecr) .
\end{equation}
Using this, it is elementary to show that,
\begin{equation}\label{eq.lagr.rot.mag.self}
L_{\rm mag}=\frac{1}{2c}\int \vecj(\vecr)\cdot \vecA(\vecr) \,\dfd
V =\frac{1}{2c}\int (\vecj_1 \cdot\vecA_1 + \vecj_2 \cdot\vecA_2
+2\vecj_1 \cdot\vecA_2) \,\dfd V ,
\end{equation}
is given by,
\begin{equation}\label{eq.lagr.rot.mag.self.expl}
L_{\rm mag}=\frac{2}{35}\frac{Q^2 R}{c^2}\vecomega^2,
\end{equation}
where, $\vecomega=\vecOmega_1-\vecOmega_2$, is the relative
angular velocity.

The response to a constant external field $\vecB$ with vector
potential $\vecA_{\rm e}=\half \vecB\times\vecr$, is given by,
\begin{equation}\label{eq.lagr.rot.mag.ext}
L_{\rm e}=\frac{1}{c}\int \vecj(\vecr)\cdot \vecA_{\rm e}(\vecr)
\,\dfd V =\frac{1}{c}\int (\vecj_1 \cdot\vecA_{\rm e} + \vecj_2
\cdot\vecA_{\rm e} ) \,\dfd V .
\end{equation}
One finds,
\begin{equation}\label{eq.lagr.rot.mag.ext.expl}
L_{\rm e}=\frac{QR^2}{10 c}\vecB \cdot \vecomega ,
\end{equation}
from straightforward calculations.


\newpage
\noindent {\large\bf Figure captions}

\begin{figure}[h]
\centering
\caption{\small Some notation for the two-sphere plasma model
treated in this article. Note the separation of the spheres is
exaggerated for clarity. It is assumed small in the relative
translation case and zero in the relative rotation case.
 \label{FIG0}}
\end{figure}
\begin{figure}[h]
\centering
\caption{\small Integration over the upper sphere, centered on
$q$, is performed by means of volume elements $\dfd V = S\dfd\rho$
that consist of the space between two concentric sphere segments
centered on $Q$ with radii $\rho$ and $\rho+\dfd\rho$. The area
$S$ of such a segment is given by $S= 2\pi \rho h$ where
$h=\rho(1-\cos\alpha)$. The cosine theorem applied to the triangle
$QqP$ gives $R^2=r^2+\rho^2-2r\rho\cos\alpha$. This gives
$\cos\alpha$ and insertion into $\dfd V$ gives $\dfd V= (\pi
\rho/r) [R^2-(r-\rho)^2]\dfd\rho$.
 \label{FIG1}}
\end{figure}
\newpage
\begin{figure}[h]
\centering
\includegraphics[width=350pt]{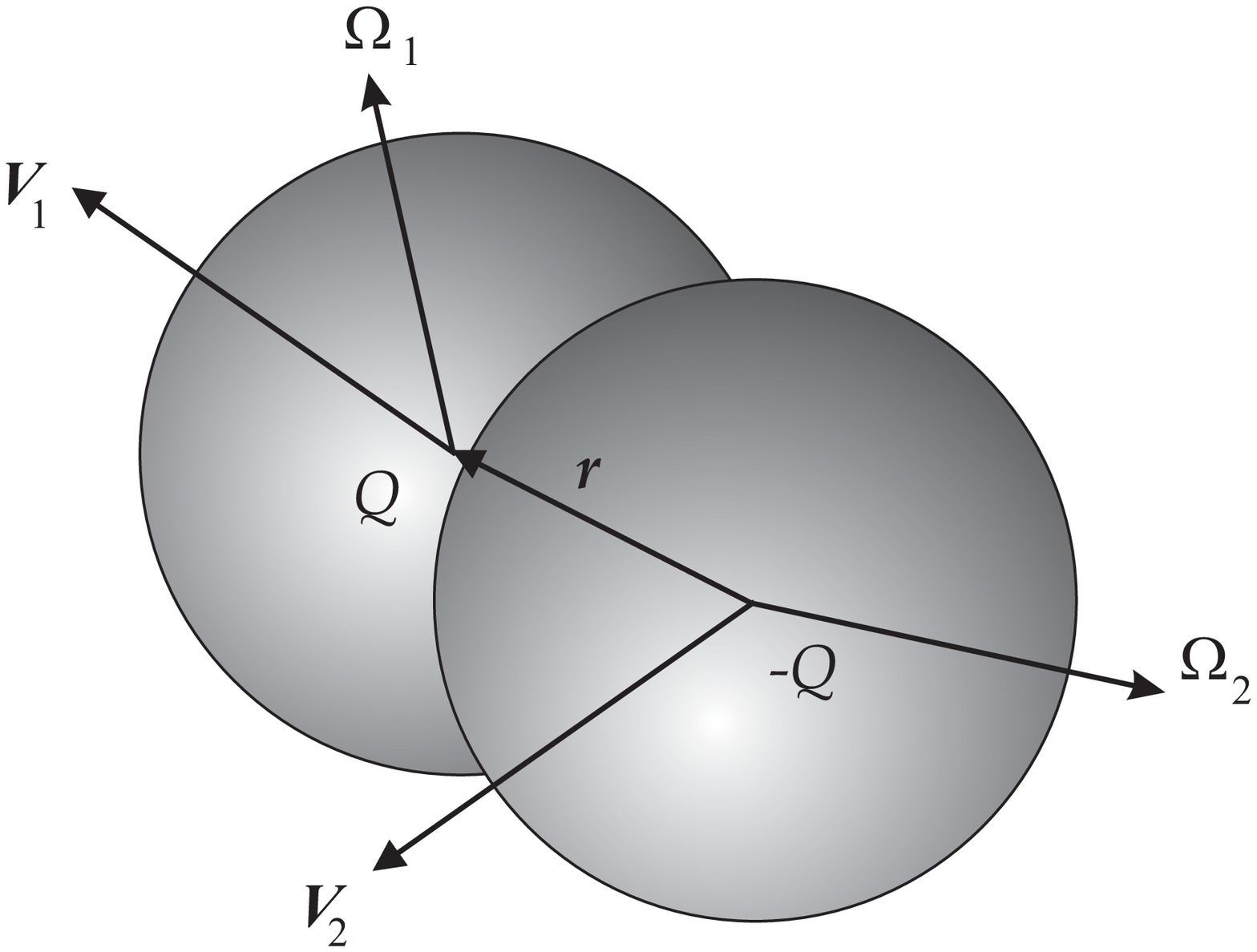}
\end{figure}
\begin{figure}[h]
\centering
\includegraphics[width=350pt]{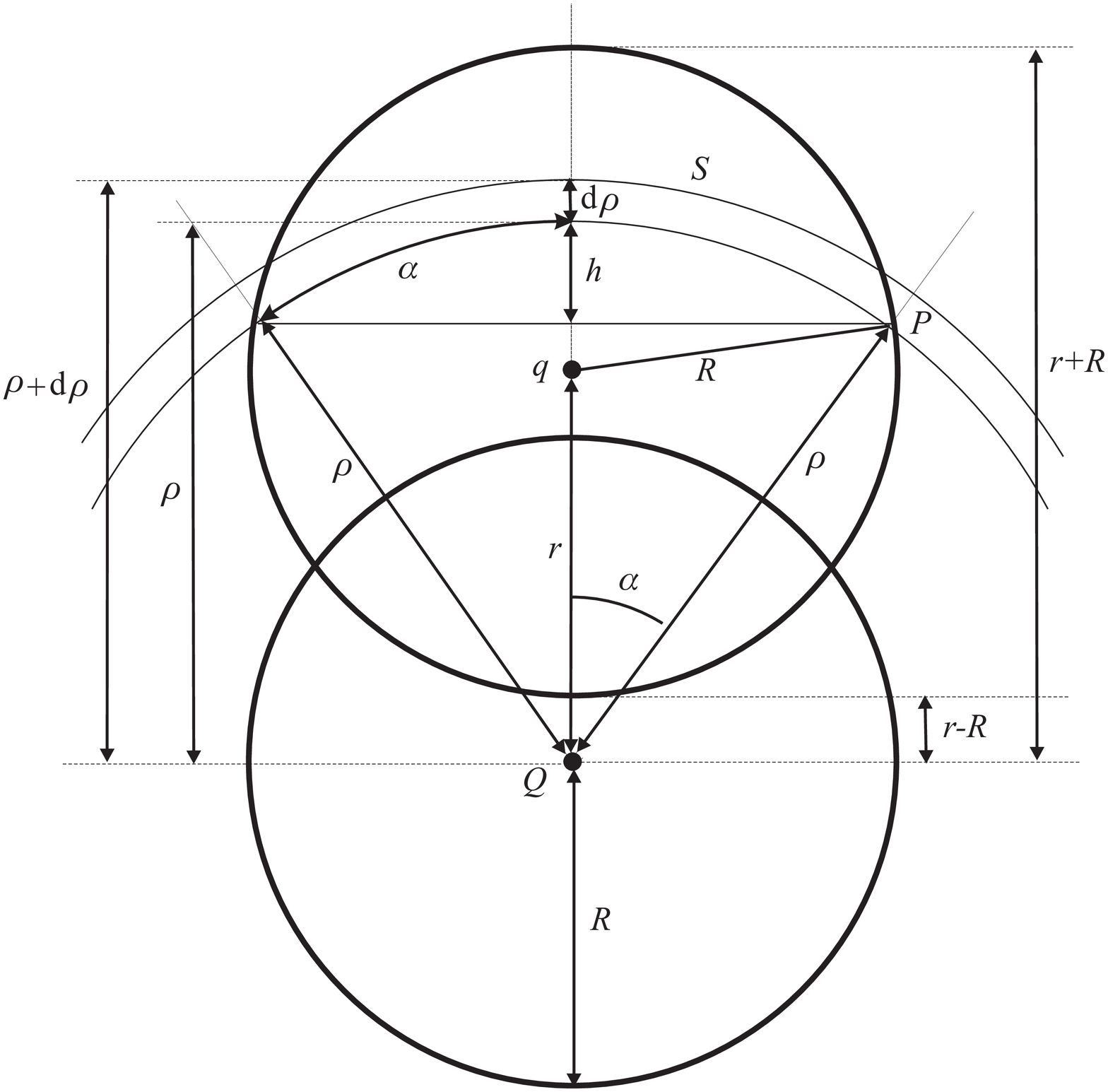}
\end{figure}
\end{document}